\documentclass{aa}  

\usepackage{graphicx}
\usepackage{txfonts}
\usepackage[colorlinks]{hyperref}
\hypersetup{
     colorlinks   = true,
     citecolor    = blue
}
\usepackage[utf8]{inputenc}
\usepackage{mathrsfs}
\usepackage{amssymb}
\usepackage{amsmath,bm}
\usepackage{physics}
\usepackage{relsize}
\usepackage{float}
\usepackage{textcomp}
\newcommand{\ts}{\textsuperscript}
\newcommand{\bxi}{\bm{\xi}}
\newcommand{\bOmega}{\bm{\Omega}}

\newcommand{\bnabla}{\bm{\nabla}}
\newcommand{\bu}{\bm{u}}
\newcommand{\br}{\bm{r}}
\newcommand{\ml}{\mathscr{L}}
\newcommand{\diffrot}{\Omega (r, \theta)}
\newcommand{\tdiffrot}{\Omega (r, \theta, t)}

\newcommand{\zdiffrot}{\Omega_0 (r, \theta)}

\newcommand{\id}{{\rm d}}
\newcommand{\rSun}{{R_{\odot}}}

\newcommand{\uth}{\bm{\hat{\theta}}}
\newcommand{\uz}{\bm{\hat{z}}}
\newcommand{\uphi}{\bm{\hat{\phi}}}
\newcommand{\ur}{\bm{\hat{r}}}
\usepackage{xcolor}

\begin{document} 

   \title{Predicting frequency changes of global-scale solar Rossby modes due to solar cycle changes in internal rotation}

   \author{C. R. Goddard
          \inst{1}
          \and
          A. C. Birch
          \inst{1}
          \and
          D. Fournier
          \inst{1}
          \and
          L. Gizon
          \inst{1,2,3}}

   \institute{Max-Planck-Institut für Sonnensystemforschung, Justus-von-Liebig-Weg 3, 37077 Göttingen, Germany\\
         \email{crgoddard10@gmail.com; gizon@mps.mpg.de}
              \and
              Institut für Astrophysik, Georg-August-Universität Göttingen, Friedrich-Hund-Platz 1, 37077 Göttingen, Germany
              \and
              Center for Space Science, NYUAD Institute, New York University Abu Dhabi,  Abu Dhabi, UAE
             }

   \date{Received 30\ts{th} May 2020}

  \abstract
   {Large-scale equatorial Rossby modes have been observed on the Sun over the last two solar cycles.
   }
   {We investigate the impact of the time-varying zonal flows on the frequencies of Rossby modes.}
   {A first-order perturbation theory approach is used to obtain an expression for the expected shift in the mode frequencies due to perturbations in the internal rotation rate. }
   {Using the time-varying rotation  from helioseismic inversions we predict the changes in Rossby mode frequencies with azimuthal orders from $m = 1$ to $m = 15$ over the last two solar cycles. 
   The peak-to-peak frequency change is  less than 1 nHz for the $m=1$ mode, grows with $m$, and reaches 25 nHz for $m = 15$.}
   {Given the observational uncertainties on mode frequencies due to the finite mode lifetimes, we find that  the predicted frequency shifts are near the limit of detectability.}
   \titlerunning{Frequency changes of  global-scale solar Rossby modes}
   \maketitle

\section{Introduction}

Rossby waves are global-scale waves of vorticity which arise due to the conservation of the radial component of the absolute vorticity for rotating fluids. They are most commonly discussed in the context of planetary atmospheres and oceans \citep[e.g.][]{vallis_2017}. 
Detections of their stellar analogues, also known as r~modes, have been proposed for Gamma Doradus stars \citep{2016A&A...593A.120V, 2018MNRAS.474.2774S, 2018arXiv181201253S}. 
Recently Rossby waves have been unambiguously detected in the near-surface flows of the Sun \citep{2018NatAs...2..568L}; they are  characterised by a robust dispersion relation which is close to that of classical Rossby waves in a uniformly rotating fluid  \cite[e.g.][]{Saio1982}. 
Follow-up studies with helioseismic methods and data have confirmed the   detection of modes with azimuthal orders $m =$~3~--~15  \citep{2019A&A...626A...3L, 2019ApJ...871L..32H, 2020A&A...635A.109H}.  The measured latitudinal eigenfunctions deviate from sectoral Legendre polynomials \citep{2020prox}. Calculations in the equatorial $\beta$-plane suggest that the effect of latitudinal differential rotation can account for the real part of the observed latitudinal eigenfunctions \citep{2020gizon}.
The observed Rossby waves can be described as 
equatorial modes  trapped by latitudinal  differential rotation. 

The theory of r~modes in the solar and stellar context has developed over the last few decades for various limiting cases \citep[e.g.][]{1981A&A....94..126P, 1998ApJ...502..961W, 2007A&A...470..815Z, 2018MNRAS.474.2774S}. 
\cite{Damiani2020} recently computed eigenmodes in slowly rotating polytropes and confirmed the $r^m$ radial dependence of the displacement eigenfunctions proposed  by \citet{1981A&A....94..126P}.

Global-scale Rossby waves are probes of the convection zone.
In this paper we predict the effect of solar cycle variations in the zonal flows on their frequencies, since these frequencies are directly related to the solar rotation rate. This work is required to interpret future measurements of time variations, in particular to disentangle perturbations caused by solar cycle changes in rotation from perturbations caused by the evolving large-scale magnetic field.

Helioseismology with data from the Michelson Doppler Imager on the Solar and Heliospheric Observatory (SOHO/MDI) \citep{1995SoPh..162..129S} and the Global Oscillation Network Group (GONG) \citep{1996Sci...272.1292H} allowed the internal rotation profile of the Sun to be measured in the convection zone.
The latitudinal differential rotation seen at the surface extends throughout the convection zone \citep[e.g.][]{1998ApJ...505..390S}. Additionally, data from the Helioseismic and Magnetic Imager on the Solar Dynamics Observatory (SDO/HMI) \citep{2012SoPh..275..229S} confirmed these findings.
All data sets revealed changes in the rotation rate that are associated with the solar cycle.
These time-varying zonal flows or torsional oscillations were first noted at the surface \citep{1980ApJ...239L..33H} and have since been shown to extend throughout most of the convection zone \citep{Vorontsov2002}, although they weaken with depth \citep[e.g.][]{2009LRSP....6....1H}. The time-varying zonal flows consist of bands of slower and faster rotation that move towards the equator at low latitudes and towards the poles at high latitudes.
These zonal flows vary significantly   between the last two solar cycles, solar cycle 23 (SC23) and solar cycle 24 (SC24) \citep{2014SoPh..289.3435K, Howe2018}.
Additionally, there is a  difference in the average rotation between these two cycles. In SC24 the convection zone rotates more quickly at low latitudes, but more slowly at high latitudes \citep[e.g.][]{2019ApJ...883...93B}. 

 Section~\ref{rot_data} describes the internal rotation data. In Section~\ref{derive_kernel} we use a first-order perturbation approach (Born approximation)  to obtain a kernel for the sensitivity of Rossby waves to variations of an arbitrary rotation profile, and the predictions via a series of steady problems are presented in Section~\ref{res}. Finally, the discussion and conclusions are given in Section~\ref{conc}. 

\section{Solar cycle variations of internal rotation}
\label{rot_data}

\begin{figure}
\centering
\includegraphics[width=9cm]{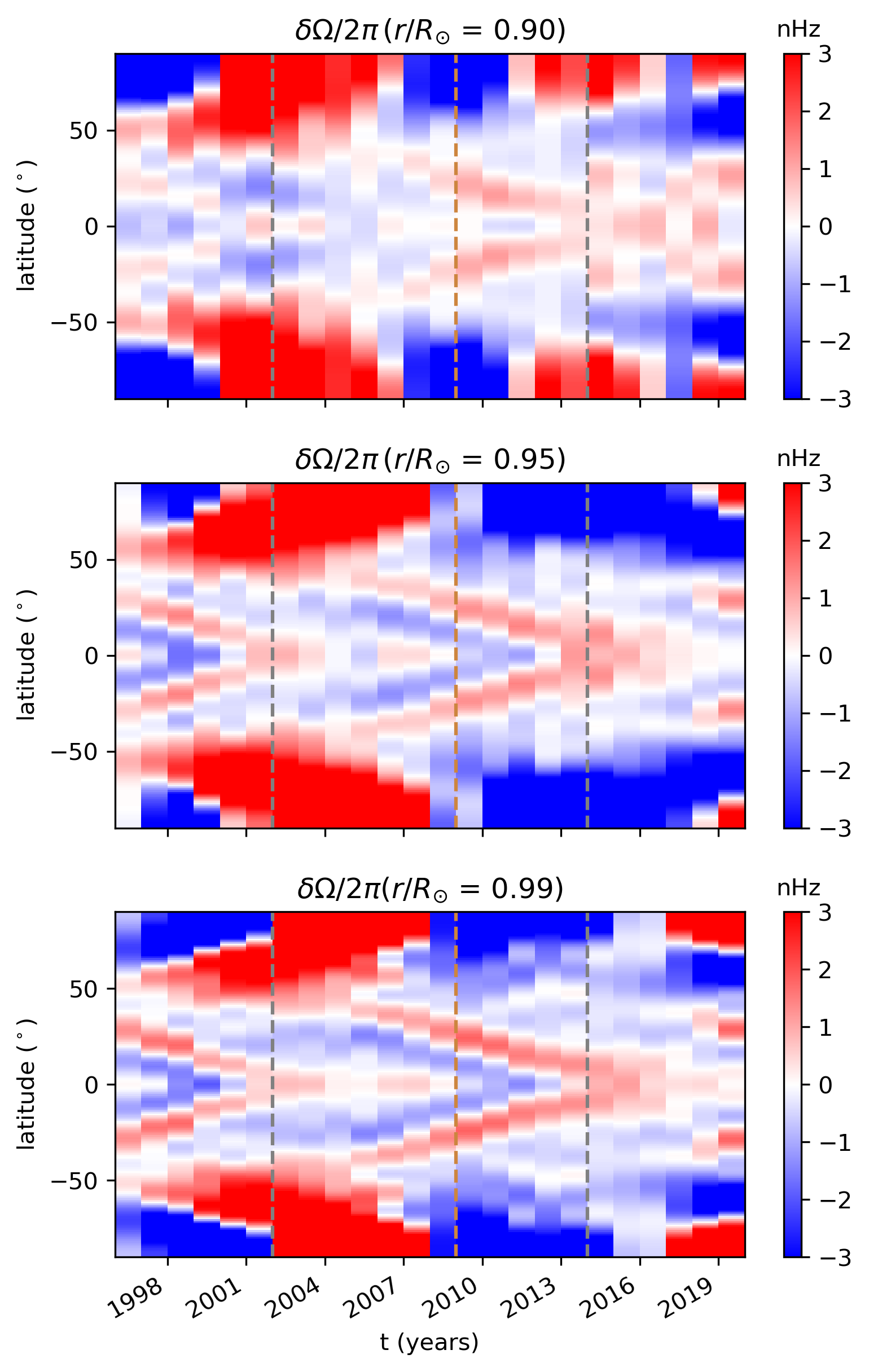}
\caption{Rotation rate residuals, $\delta\Omega(r,\theta,t)/2\pi$, at $r$ = 0.9, 0.95, and 0.99 $R_{\odot}$ for 1996-Jun-06 to 2019-Jun-30. The data is binned for each calendar year, and   the colour scale is saturated at $\pm$ 3 nHz. The vertical dashed grey lines denote the positions of the maxima of solar cycles 23 and 24  (2002 and 2014). The vertical dashed brown line indicates the minimum between the two cycles (2009).}
 \label{fig:rot_res}
\end{figure}

The rotation rate data used to prescribe the zonal flows are the 2D regularised least-squares (RLS) global helioseismology inversions for solar rotation from MDI  and HMI \citep{2018SoPh..293...29L}. The data used here cover the period from 1996 June 06 to 2019\ June 30, with a cadence of 72 days. The HMI data are used during the period of overlap between MDI and HMI. We work in spherical-polar coordinates, where $r$ is the radial coordinate, $\phi$ is longitude,  and $\theta$ is the co-latitude. A reference rotation profile, $\overline{\Omega}(r,\theta)$, is taken to be the unweighted mean rotation profile over the 23-year  time interval. Variations about this mean profile over time are thus
\begin{equation}
    \delta\Omega(r, \theta, t) = \Omega(r, \theta, t) - \overline{\Omega}(r,\theta).
    \label{eq:rot_res}
\end{equation}The few missing time steps are dealt with via linear interpolation. The rotation residual is then binned by taking the unweighted mean for each calendar year.

Figure~\ref{fig:rot_res} shows the binned rotation rate residual, $\delta\Omega(r, \theta, t)$, at $r/R_\odot$ = 0.90, 0.95, and~0.99 for the specified time window. The zonal flows and high latitude variations associated with the 11-year solar cycle are seen alongside the asymmetry between SC23 and SC24. As mentioned previously, in the convection zone SC24 has higher rotation rates at low latitudes, but lower rotation rates at high latitudes. This asymmetry is not due to the inconsistencies between the MDI and HMI data, and is clearly seen in GONG data \citep{2019ApJ...883...93B}. 

\section{Perturbation to Rossby mode frequencies due to variations in rotation}
\label{derive_kernel}

We utilise first-order perturbation theory  to obtain a kernel for the sensitivity of Rossby wave frequencies to small variations in the  rotation profile.  This method is commonly  used to obtain sensitivity kernels for the effect of rotation on  p modes \cite[e.g.][]{1990MNRAS.242...25G, 2009LRSP....6....1H, 2016LRSP...13....2B}. 

\subsection{Oscillation equation}
We start from the linearised equation of motion for the displacement vector $\bxi$, written in an inertial frame \citep{1967MNRAS.136..293L, 1989nos..book.....U}. Assuming a steady rotation rate $\Omega(r,\theta)$ (the wave period is  shorter than the timescale of the evolution of rotation) and a displacement vector proportional to  $\exp(-i\omega t)$, we have 
\begin{equation}
\omega^2 \bxi + \omega B (\bxi) - T (\bxi) = V(\bxi) + P(\bxi),
\label{eq.orig}
\end{equation}
with
\begin{eqnarray}
B(\bxi) &=& 2 i (\bu \cdot \bnabla)\bxi , \nonumber\\
T(\bxi) &=& (\bu \cdot \bnabla)^2\bxi , \nonumber\\
V(\bxi) &=&  - d(\bnabla \psi) ,\nonumber\\
P(\bxi) &=& d \left( \frac{1}{\rho} \bnabla p \right)  , \nonumber 
\end{eqnarray}
where  
\begin{equation}
    \bu = \bOmega\times \br = r \sin{\theta}\ \diffrot \, \uphi 
\end{equation}
is the rotational velocity, $\br$ is the radial position vector, $p$ is pressure, $\psi$ is the gravitational potential, and $d$ is the Lagrangian change operator (e.g.  $d p = p' + \bxi\cdot\bnabla p_0$). Frictional and electromagnetic forces are not considered. 

The following relation holds if we assume the azimuthal dependence of the displacement to be proportional to  $\exp(i m \phi)$, where $m$ is the azimuthal order of a solar Rossby mode:
\begin{equation}
    (\bu\cdot\bnabla) \bxi = i m \Omega \bxi + \bOmega \times\bxi .
\end{equation}
Using this relationship we can rewrite the  operators that depend on $\bu$ as
\begin{eqnarray}
    B (\bxi) &=& - 2 m \Omega \bxi + 2 i \bOmega\times\bxi , \\
    T (\bxi) &=& -m^2 \Omega^2 \bxi + 2 i m \Omega \bOmega\times\bxi + \bOmega\times(\bOmega\times\bxi) .
\end{eqnarray}
Substituting these into Eq.~(\ref{eq.orig}), and defining $\ml(\bxi) = V(\bxi) + P(\bxi)$, gives
\begin{equation}
    (\omega- m \Omega)^2 \bxi + 2 i (\omega-m \Omega) \bOmega\times\bxi - \bOmega\times(\bOmega\times\bxi) - \ml (\bxi) = 0.
\label{inert_eom}
\end{equation}


\begin{figure*}
\centering
\includegraphics[width=18.5cm]{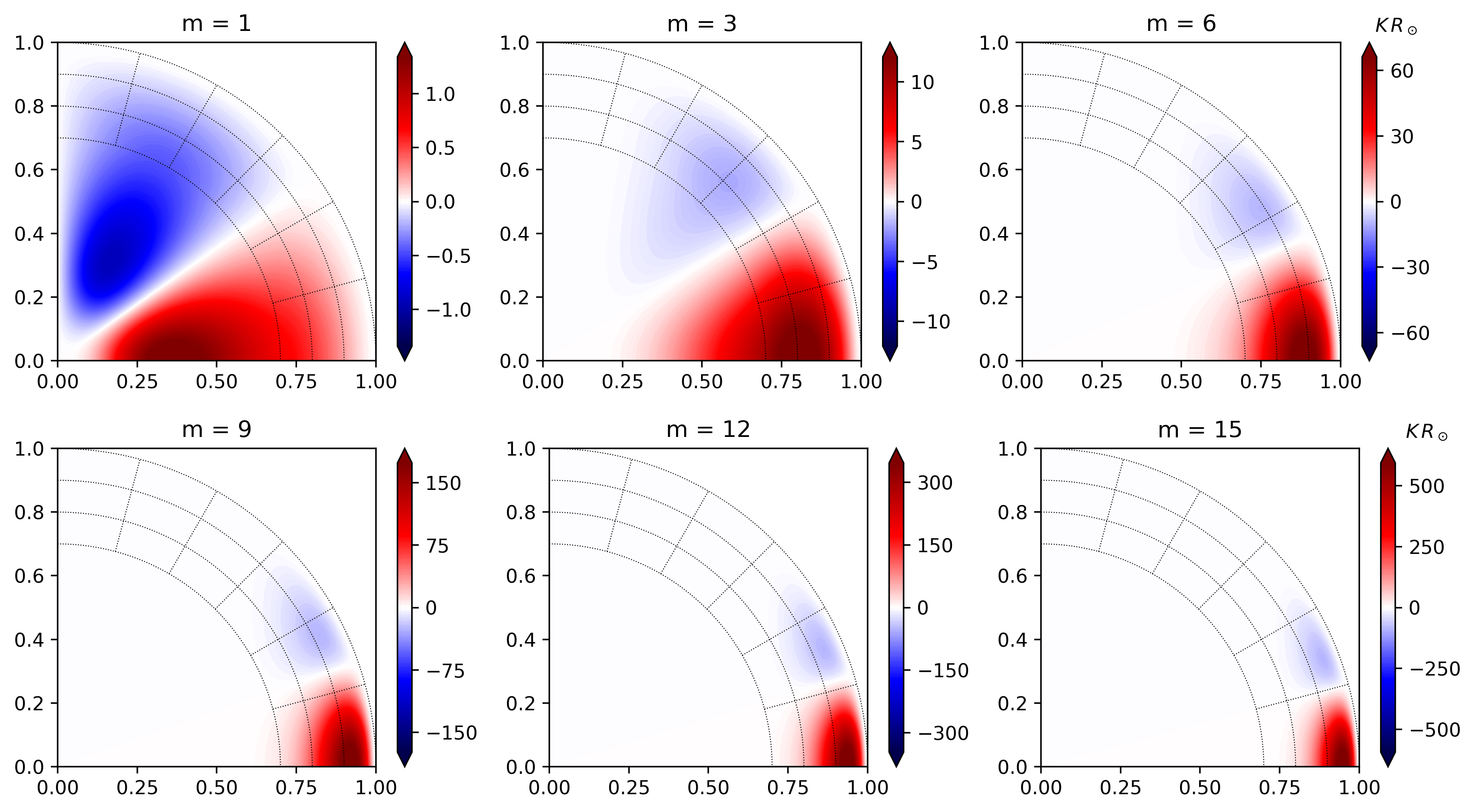}
\caption{Kernels for the sensitivity of sectoral Rossby waves to variations in rotation rate for azimuthal orders $m$ = 1, 3, 6, 9, 12, and 15 (from left to right, top to bottom). 
The different panels show the kernels in the meridional plane and the kernels are symmetric about the equator. The vertical axis is the rotation axis. The spatial grid is in units of   $R_\odot$, and  the grided region denotes the convection zone.
}
 \label{ker_plots}
\end{figure*}

\subsection{First-order oscillation equation}
We now perturb Eq.~(\ref{inert_eom}) around an arbitrary  reference rotation rate $\zdiffrot$ using
\begin{eqnarray}
    \diffrot &=& \Omega_0(r, \theta) + \delta\diffrot , \nonumber\\
    \omega &=& \omega_0 + \delta\omega , \nonumber\\
    \bxi &=& \bxi_0 + \delta\bxi , \nonumber\\
    \ml &=& \ml_0 + \delta\ml .
\end{eqnarray}
The zero-order eigenvalue problem  for $(\omega_0, \bxi_0)$  is 
\begin{equation}
    (\omega_0- m \Omega_0)^2 \bxi_0 + 2 i (\omega_0-m \Omega_0) \bOmega_0\times\bxi_0 - \bOmega_0\times(\bOmega_0\times\bxi_0) - \ml_0 (\bxi_0) = 0  .
\label{zero_order_eom}
\end{equation}
The first-order eigenvalue problem for $(\delta \omega, \delta\bxi)$ is 
\begin{align}
    &(\omega_0-m\Omega_0)^2\delta\bxi + 2(\omega_0-m\Omega_0)(\delta\omega-m\delta\Omega)\bxi_0 \nonumber\\ 
    &+ 2 i (\omega_0-m\Omega_0) \Omega_0\times\delta\bxi + 2 i (\omega_0-m\Omega_0) \delta \bOmega \times \bxi_0 \nonumber\\
    &+ 2 i (\delta\omega-m\delta\Omega) \Omega_0\times\bxi_0 - \delta\bOmega\times(\bOmega_0\times\bxi_0) - \bOmega_0\times(\delta\bOmega\times\bxi_0) \nonumber \\
    &- \bOmega_0\times(\bOmega_0\times\delta\bxi) - \ml_0 (\delta\bxi) -\delta \ml (\bxi_0)=0.
\label{fin_eom}
\end{align}

\subsection{First-order frequency shift}
We define an inner product of two functions as ${\langle \bm{\eta}, \bxi\rangle = \int_V \bm{\eta}^*\cdot\bxi\  \rho \id V}$, where  $V$ is the  volume of the star, the asterisk denotes the complex conjugate, and $\rho(r)$ is the radial density profile.

Dotting Eq.~(\ref{fin_eom}) with $\bxi^*_0$ on the left, integrating over $V$, and utilising the self-adjointness of the operators \citep[cf.][]{1967MNRAS.136..293L} gives 
\begin{align}
    & 2\langle \bxi_0,(\omega_0-m\Omega_0)(\delta\omega-m\delta\Omega)\bxi_0\rangle + 2 i \langle \bxi_0, (\omega_0-m\Omega_0)\delta\Omega\ \uz \times \bxi_0 \rangle \nonumber \\
     &  + 2 i \langle\bxi_0, (\delta\omega-m\delta\Omega) \Omega_0\ \uz\times\bxi_0 \rangle 
     - 2 \langle\bxi_0, \delta\Omega\ \Omega_0\ \uz\times(\uz\times\bxi_0) \rangle 
     \nonumber \\
     & - \langle\bxi_0,  \delta \ml (\bxi_0)\rangle  =0,
\label{fin_pert_eom}
\end{align}
where  $\uz$ is the unit vector along the rotation axis, $\uz = \ur\cos{\theta}  - \uth \sin{\theta} $. Using this expression the frequency shift due to variations in an arbitrary rotation profile may be calculated from the unperturbed (zero-order) displacement eigenfunctions. Rearranging the terms that include $\delta\omega$, and defining $\mathscr{S}(\bxi)$ to contain the centrifugal and $\delta\ml$ terms, 
\begin{equation}
{\mathscr{S}(\bxi_0)= \delta\Omega\ \Omega_0\ \uz\times(\uz\times\bxi_0) +\frac{1}{2}\delta \ml\ (\bxi_0)}, 
\end{equation}
we have
\begin{equation}
\begin{split}
    \delta\omega = \frac{ \langle \bxi_0, \delta \Omega\ \mathscr{N}(\bxi_0) + \mathscr{S}(\bxi_0)\rangle}{\langle  \bxi_{0} ,   \bm{\mathscr{D}}(\bxi_0)\rangle}  \; ,
\end{split}
\label{freq_shift_sim}
\end{equation}
with
\begin{align}
    \mathscr{N}(\bxi_0) = &  - i (\omega_0-2 m\Omega_0) (\uz \times \bxi_0) + m (\omega_0 - m\Omega_0)\bxi_0,
   \\
    \mathscr{D}(\bxi_0) =&  (\omega_0-m\Omega_0) \bxi_0 + i\Omega_0\ \uz\times\bxi_0 \; .
\end{align}
We note that the denominator is more complicated than in the standard p-mode case with a non-rotating reference model.
For the rest of the paper, we   neglect the term $\mathscr{S}(\bxi)$ in Eq.~(\ref{freq_shift_sim}), which corresponds to neglecting the perturbation of the centrifugal acceleration and the perturbation to the pressure and gravity terms due to the change in rotation $\delta\Omega$.

\subsection{Rossby wave eigenfunction and resulting kernel}
\label{ker}

From here forwards we consider reference eigenfunctions $\bxi_0$ which correspond to incompressible global Rossby waves for the case of slow uniform rotation. In this case deviations of the solar structure (i.e. oblateness) from variations in rotation have a negligible impact on the eigenfrequencies \citep{Damiani2020}. The relevant zero-order equation of motion, Eq.~(\ref{zero_order_eom}) without the centrifugal (third) term, is now
\begin{equation}
    (\omega_0- m \Omega_0)^2 \bxi_0 + 2 i (\omega_0-m \Omega_0) \bOmega_0\times\bxi_0 - \ml_0 (\bxi_0) = 0.
\label{zero_order_eom_ros}
\end{equation}
This is the inertial frame equivalent of the oscillation equation given in \cite{1981A&A....94..126P} and \cite{Damiani2020}, under the Cowling approximation. 

For the case of slow uniform rotation, the eigenfunction of the Rossby mode with azimuthal wavenumber $m$ is of the form
\begin{eqnarray}
\bxi_0 &=& \bnabla_h\times\left( \bm{r} R(r) L(\theta)e^{i(m\phi-\omega_0 t)}  \right) \nonumber\\
&=& R(r)   \left( {\uth \frac{imL(\theta)}{\sin\theta}   - \uphi} L'(\theta) \right)e^{i(m\phi-\omega_0 t)} , 
\label{ros_sol}
\end{eqnarray}
where $\bnabla_h$ is the horizontal gradient, $R(r)$ and $L(\theta)$ are  functions to be specified for each mode, and $L'(\theta) = dL/d\theta$.

\begin{figure}
\centering
\includegraphics[width=9cm]{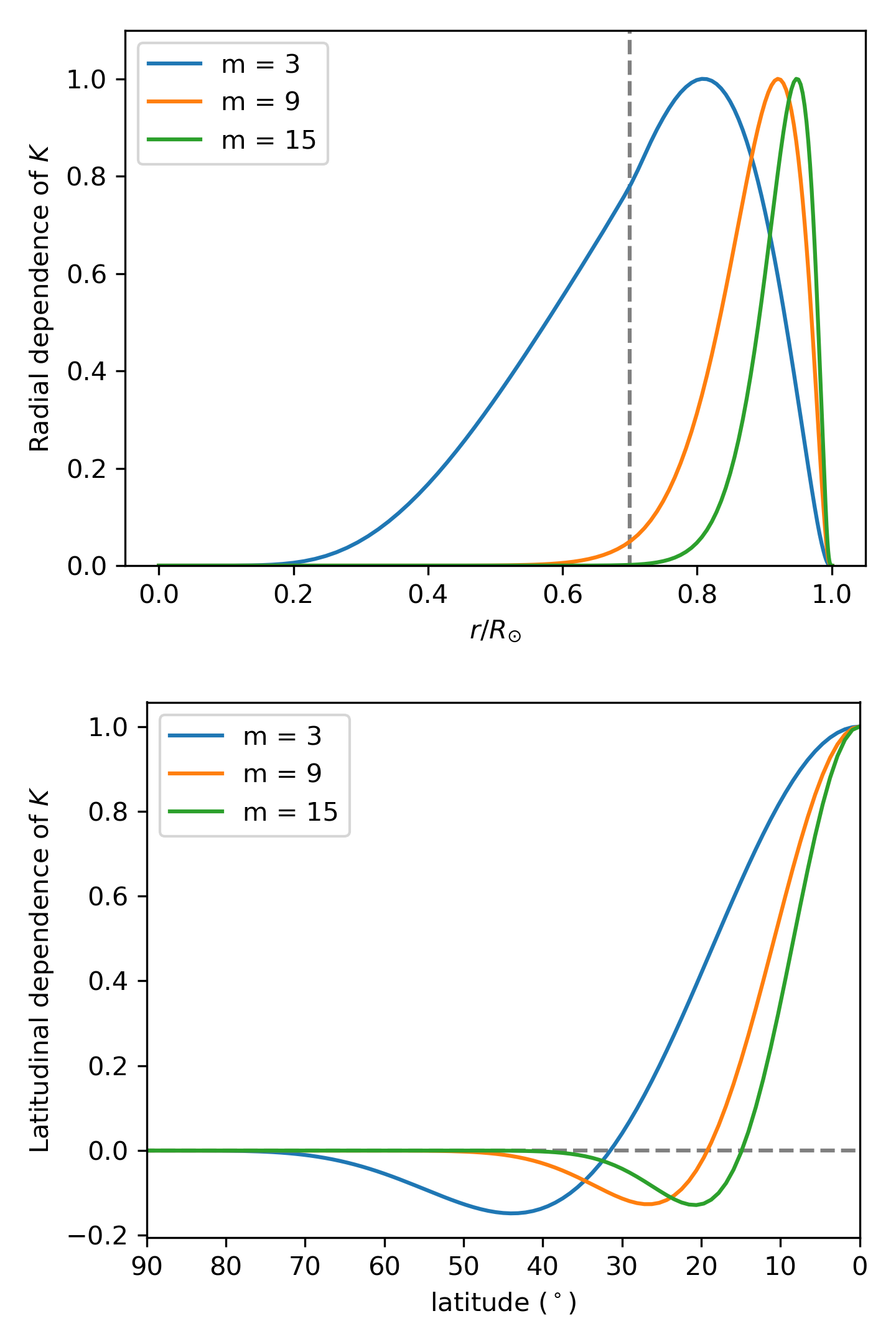}
\caption{Radial (top) and latitudinal (bottom) dependences of the kernel for sectoral Rossby modes with  azimuthal orders $m$ = 3, 9, and 15, normalised by their maxima. The vertical dashed line in the top panel denotes the position of the base of the convection zone at $r/R_\odot = 0.7$.}
 \label{ker_cuts}
\end{figure}

%
The frequency shift to the unperturbed eigenfrequency  $\omega_0$ of mode $m$ due to a change in rotation is given by
\begin{equation}
    \delta \omega = \int_0^\pi\int_0^\rSun    K (r,\theta)\ \delta\Omega(r,\theta) \,  \id r \id\theta,
\label{freq_shift_temp}
\end{equation}
where the sensitivity kernels $K$ are given by
\begin{equation}
    K (r,\theta) = \frac{\bxi^{*}_{0} \cdot  \bm{\mathscr{N}}(\bxi_0) \;  \rho r^2 \sin\theta}{\int_0^\pi\int_0^\rSun \bxi^{*}_{0} \cdot  \bm{\mathscr{D}}(\bxi_0) \; \rho r^2\sin\theta \,  \id r \id\theta} .
\label{ker_comp}
\end{equation}

For the displacement eigenfunction given in Eq.~(\ref{ros_sol}) and a real-valued $L(\theta)$, we have
\begin{align}
    \bxi_0^* \cdot \mathscr{N}(\bxi_0) =  & m R^2(r)  \left[ \left(4m\Omega_0(r, \theta)-2\omega_0\right)\left(\frac{L(\theta)L'(\theta)\cos\theta}{\sin\theta}\right) \right. \nonumber \\
    & + \left. \left(\omega_0 - m\zdiffrot\right) \left( \frac{m^2 L^2(\theta)}{\sin^2\theta} + L'(\theta)^2 \right) \right] 
\end{align}
and 
\begin{align}
    \bxi_0^* \cdot \mathscr{D}(\bxi_0) = &  R^2(r)\left[  2m\zdiffrot \left(\frac{L(\theta)L'(\theta) \cos\theta}{\sin\theta}\right)\nonumber \right. \nonumber \\
    & \left.  + (\omega_0-m\zdiffrot)\left(\frac{m^2 L^2(\theta)}{\sin^2\theta} + L'(\theta)^2\right) \right].
\end{align}

Now we choose specific radial and latitudinal dependences for the eigenfunction given by Eq.~(\ref{ros_sol}), along with a corresponding reference eigenfrequency. 
Considering the case of uniform slow rotation
\begin{equation}
    \Omega_0(r,\theta) = \Omega_0 = \textrm{ const} ,
\end{equation}
we take a sectoral spherical harmonic ($\ell=m$) solution with no radial node:
\begin{eqnarray}
    R(r) &=&  r^{m}\\
    L(\theta) &=& (\sin\theta)^m \\
    \omega_0 &=& m \Omega_0 - \frac{2 \Omega_0}{m+1}.
\end{eqnarray}
This simple form for the radial eigenfunctions is suggested by the study of r modes in  slowly (and uniformly) rotating polytropes in \citet{Damiani2020}. Corrections may be needed to account for strong radial gradients in solar rotation and for the sub-adiabatic stratification below the convection zone; however, this is beyond the scope of this paper. We note that the kinetic energy density of most of the modes we consider here (m $\geq$ 4) is largely restricted to the convection zone.

For this case, the kernel simplifies to
\begin{equation}
    K (r,\theta) = \frac{\rho r^{2(m+1)} (\sin\theta)^{2m-1} \left[m - (m^2+2)\cos^2\theta \right] }{\int_0^\rSun  r^{2(m+1)} \rho \id r  \int_0^\pi (\sin\theta)^{2m-1}  (1 - m\cos^2\theta) \,   \id\theta} \; .
    \label{simp_ker}
\end{equation}
This kernel does not depend on the uniform rotation rate $\Omega_0$ from the zero-order model. Appendix~\ref{app:const_test} shows that this kernel predicts the correct answer for the case of a perturbation $\delta\Omega$ that is constant.

Figure~\ref{ker_plots} shows the kernel calculated with the smoothed solar-like density profile from \citet{2017A&A...600A..35G}. Figure~\ref{ker_cuts} shows normalised latitudinal and radial cuts for azimuthal orders $m$ = 3, 9, and 15. The frequency shifts are mostly sensitive to perturbations of the rotation rate in the low latitude convection zone for $m \geq$ 3. The negative higher latitude component of the kernel reduces in amplitude with increasing azimuthal order. The detected solar Rossby waves currently span $m$ = 3 -- 15, and hence the plots are restricted to $m \leq$ 15. Appendix~\ref{app:lat_form} discusses the latitudinal dependence of the kernel in more detail.

\section{Results}
\label{res}

\begin{figure}
\centering
\includegraphics[width=9cm]{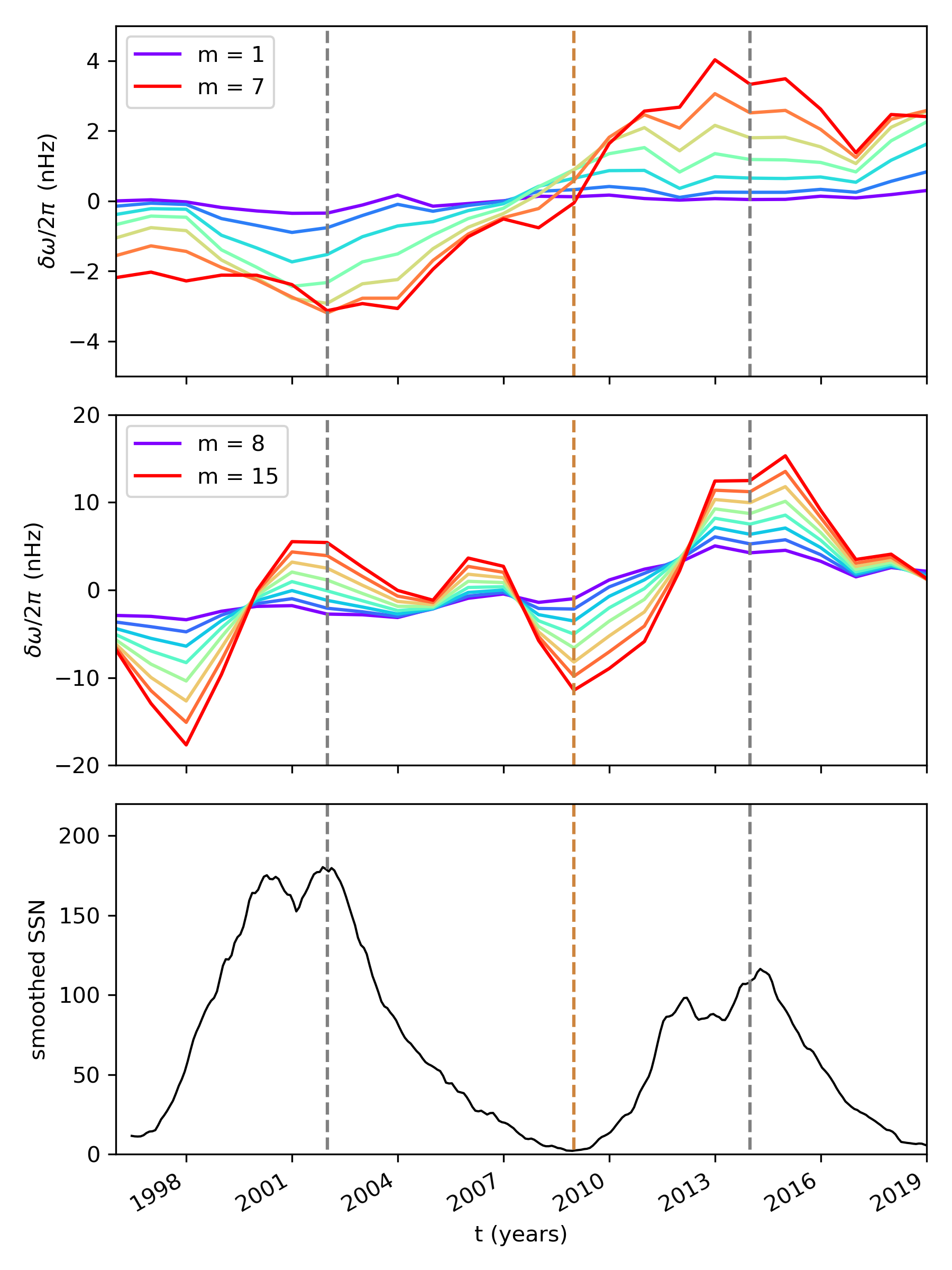}
\caption{Rossby mode frequency-shift time series due to variation in the solar rotation profile, plotted for $m$ from 1 to 8 (top), and 9 to 15 (middle), calculated for each calendar year. The vertical grey dashed  lines denote the positions of the maxima of solar cycles 23 and 24, to the nearest year. The vertical brown dashed line indicates the minimum between the two cycles. The smoothed sunspot number is plotted in the bottom panel.}
\label{freq_shift}
\end{figure}

\begin{figure}
\centering
\includegraphics[width=9cm]{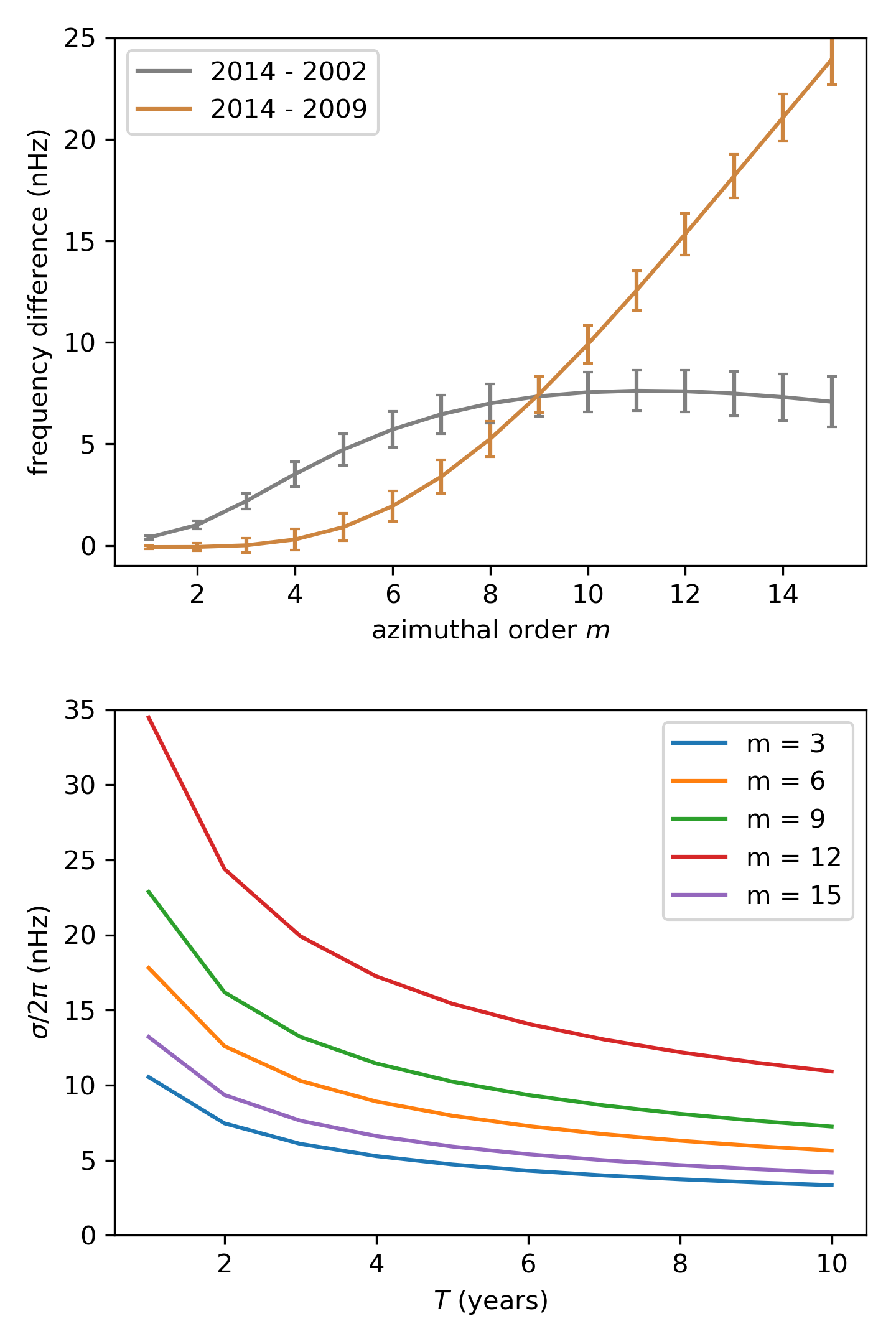}
\caption{\textit{Top:} Predicted difference in the Rossby mode frequencies between the two solar maxima (grey curve, 2014 minus 2002) and between the maximum of cycle~24 and the preceding solar minimum (brown curve, 2014 minus 2009).
The error bars are propagated errors from the scatter in the solar rotation measurements.
\textit{Bottom:} 
Estimated errors in the Rossby mode frequencies due to the finite mode lifetimes as a function of observation time $T$. 
These errors can be used to determine whether the predicted frequency shifts from the top panel are detectable.
}
\label{freq_shift_range}
\end{figure}

\subsection{Predicted frequency shifts over the last two solar cycles}

Here we compute the Rossby wave frequency shifts $\delta\omega$ for each mode $m$, which  would be expected from the 
mean-subtracted rotation residuals, $\delta\tdiffrot$,
defined in Eq.~(\ref{eq:rot_res}).  The rotation residuals are binned by calendar year, as shown in Fig.~\ref{fig:rot_res}. We do not consider the time-independent frequency shift for each $m$ caused by the difference between our zero-order model (uniform rotation, $\Omega_0$) and the time-averaged differential rotation of the Sun ($\overline{\Omega}(r,\theta)$) used to define $\delta\tdiffrot$.

The frequency-shift time series for each azimuthal order were calculated from the binned rotation residuals and Eqs.~(\ref{freq_shift_temp}) and~(\ref{simp_ker}) using trapezoidal rule integration. The perturbation to the rotation profile $\delta\tdiffrot$ is set to zero in the radiative zone (${r/R_{\odot} > 0.7}$), due to the inaccuracies in the inversions there and the lack of zonal flow signatures at these depths. 

Figure~\ref{freq_shift} shows the obtained frequency shifts separately for low  and high azimuthal orders, $1 \leq m \leq 7$ and $8 \leq m \leq 15$, respectively.  For $m=1$ no discernible trend is seen as this mode has little sensitivity to the strongest zonal flows in the upper convection zone. From $m=2$ to $m=7$ an apparent cycle with a period of $>$~22 years is observed: the frequency shift is negative for SC23 and positive for SC24. For $m>7$ a double-peaked trend is seen, similar to the 11-year Sunspot cycle, but with the frequency during SC24  systematically higher than in SC23. This is a superposition of the 11-year cycle of the zonal flows with the asymmetry of the two cycles (as seen for low $m$).   Appendix~\ref{app:ker_int} discusses the relationship between the frequency shift and the rotation residuals in more detail. Appendix~\ref{app:beta} shows a comparison to frequency shifts computed from direct numerical calculations in the $\beta$-plane \citep{2020gizon}.

As one simple way to quantify these trends, we calculate the change in frequencies between two times for each $m$. The upper panel of Fig.~\ref{freq_shift_range} shows the difference in the frequencies between 2014 and 2002 (two solar maxima) and between 2014 and 2009 (maximum and minimum). 
The difference between maximum and minimum highlights the 11-year component, and continues to increase for high $m$, reaching 23~nHz for $m=15$. 
The difference between the cycle maxima highlights the asymmetry between the two cycles, which saturates for $m > 8$ at about 7~nHz.

\subsection{Predicted measurement uncertainties due to measured finite mode lifetimes}

Let us denote by $\sigma$ the uncertainty on the measured frequency $\omega$ of a Rossby mode $m$ due to the finite mode lifetimes. 
Under the assumption that Rossby modes are stochastically excited, 
the uncertainties $\sigma$ scale as $T^{-1/2}$, where $T$ is the observation time.
Specifically, we have ${\sigma = (\Gamma/2T)^{1/2}}$, 
where $\Gamma$ is the mode linewidth in frequency space
\citep{1992ApJ...387..712L}. 

We used  the measurements of the mode linewidths  from \cite{2019A&A...626A...3L}  to calculate the measurement uncertainties $\sigma$ for various values  $T$. 
The estimate of the uncertainties in the r mode frequencies depends on the measured lifetimes. We note that the various methods of helioseismology and granulation tracking give essentially the same answers for the mode linewidths \citep{2020A&A...635A.109H}.
The bottom panel of Fig.~\ref{freq_shift_range} shows this for several values of $m$. 
We note the strong dependence on $m$. 
For some azimuthal orders and $T=3$ years, for example, the uncertainties are smaller than the predicted solar cycle frequency shifts.
Thus, the detection of the Rossby mode frequency shifts due to the time-varying zonal flows may be possible.
However, as this is near the limit of detectability some averaging over several  values of $m$ may be required to reduce noise.

\section{Conclusion}
\label{conc}

We used first-order perturbation theory to obtain a kernel for the frequency shift of solar Rossby waves caused by perturbations in the rotation profile. We used these kernels together with solar torsional oscillation data to predict the time varying component of the frequency shift for solar Rossby waves. The predicted frequency shifts over the last two solar cycles have a strong dependence on the azimuthal order $m$. Modes with $m \leq 7$ are mainly sensitive to the asymmetry between the two cycles, while those with $m > 7$ are mainly sensitive to the 11-year  cycle of the zonal flows. The predicted frequency difference between cycle maximum and minimum reaches 23~nHz for $m=15$. 

Several  approximations and simplifications were made in the construction of the kernel, and in the chosen latitudinal and radial dependence for the sectoral Rossby waves. These are adequate for the problem at hand, but may require further development once the frequency shifts have been observed.  For example, it might be useful to consider eigenfunctions from a differentially rotating reference model. In future work we intend to investigate potential solar cycle changes in the amplitudes and the lifetimes of the modes, as well as potential changes in the eigenfunctions due to the torsional oscillations or the magnetic field.

The predicted shift of the mode frequencies due to variation in the internal rotation over the last two solar cycles is near the limit of detection. 
A detection of this effect would help further constrain the physics of solar Rossby modes. 
Should there be much larger solar cycle variations in the mode frequencies, they will have to be attributed to the effect of the solar magnetic field in the convection zone.


\begin{acknowledgements}
We thank Zhi-Chao Liang and Chris~S. Hanson for useful discussions.
Author contributions: LG proposed the research. CRG performed the kernel calculations and the data analysis. ACB and LG contributed to the kernel calculations. DF performed the $\beta$-plane calculations in Appendix \ref{app:beta}. CRG wrote the draft paper and all authors contributed to the final manuscript. 
This research is supported  by ERC Synergy grant WHOLE SUN \#810218. The data were processed at the German Data Center for SDO, funded by the German Aerospace Center (DLR). LG acknowledges partial support from the NYU Abu Dhabi Center for Space Science under grant G1502. 
The HMI and MDI helioseismology data products are used courtesy of the JSOC team. NumPy \citep{2011CSE....13b..22V} and matplotlib \citep{2007CSE.....9...90H} were used in this research. 
\end{acknowledgements}

\bibliographystyle{aa} 
\bibliography{ref.bib} 

\begin{appendix} 

\section{Case $\delta\Omega = \textrm{const}$}
\label{app:const_test}

The frequency shift in any reference frame for the described sectoral Rossby modes caused by a change in the rotation rate is
\begin{equation}
  \delta \omega = m\delta\Omega - \frac{2\ \delta\Omega}{m+1}.
\label{fshift_const}
\end{equation}
The kernel given should reduce to this result in the case of uniform rotation  ($\Omega_0={\rm const}$)  with a uniform perturbation ($\delta\Omega={\rm const}$). This has been shown via numerical integration, once converged the resulting frequency shifts agree with Eq.~(\ref{fshift_const}) to machine precision.

The kernel may also be integrated directly to obtain the frequency shift for the case of uniform rotation and perturbation.
Defining the frequency shift as
\begin{equation}
    \delta \omega =  \delta\Omega \int_0^\pi\int_0^\rSun    K (r,\theta)\,  \id r \id\theta
\end{equation}
and taking the kernel given in Eq.~(\ref{simp_ker}), the radial component cancels from top and bottom giving
\begin{equation}
    \delta \omega =   \frac{\delta\Omega \int_0^\pi(\sin\theta)^{(2m-1)} \left[m - (m^2+2)\cos^2\theta \id\theta \right] }{\int_0^\pi(\sin\theta)^{(2m-1)} \left[1 - m\cos^2\theta \right] \,  \id\theta}
.\end{equation}
Utilising the identity ${I_p=\int_0^\pi (\sin\theta)^p d\theta=  [(p-1)/p] I_{p-2} }$, we  obtain
\begin{equation}
\delta\omega = \frac{ (m^2 + m -2)}{m+1}\delta\Omega = m\delta\Omega - \frac{2\ \delta\Omega}{m+1},
\end{equation}
which is the expected result.

\section{Latitudinal form of the kernel}
\label{app:lat_form}

\begin{figure}
\centering
\includegraphics[width=9cm]{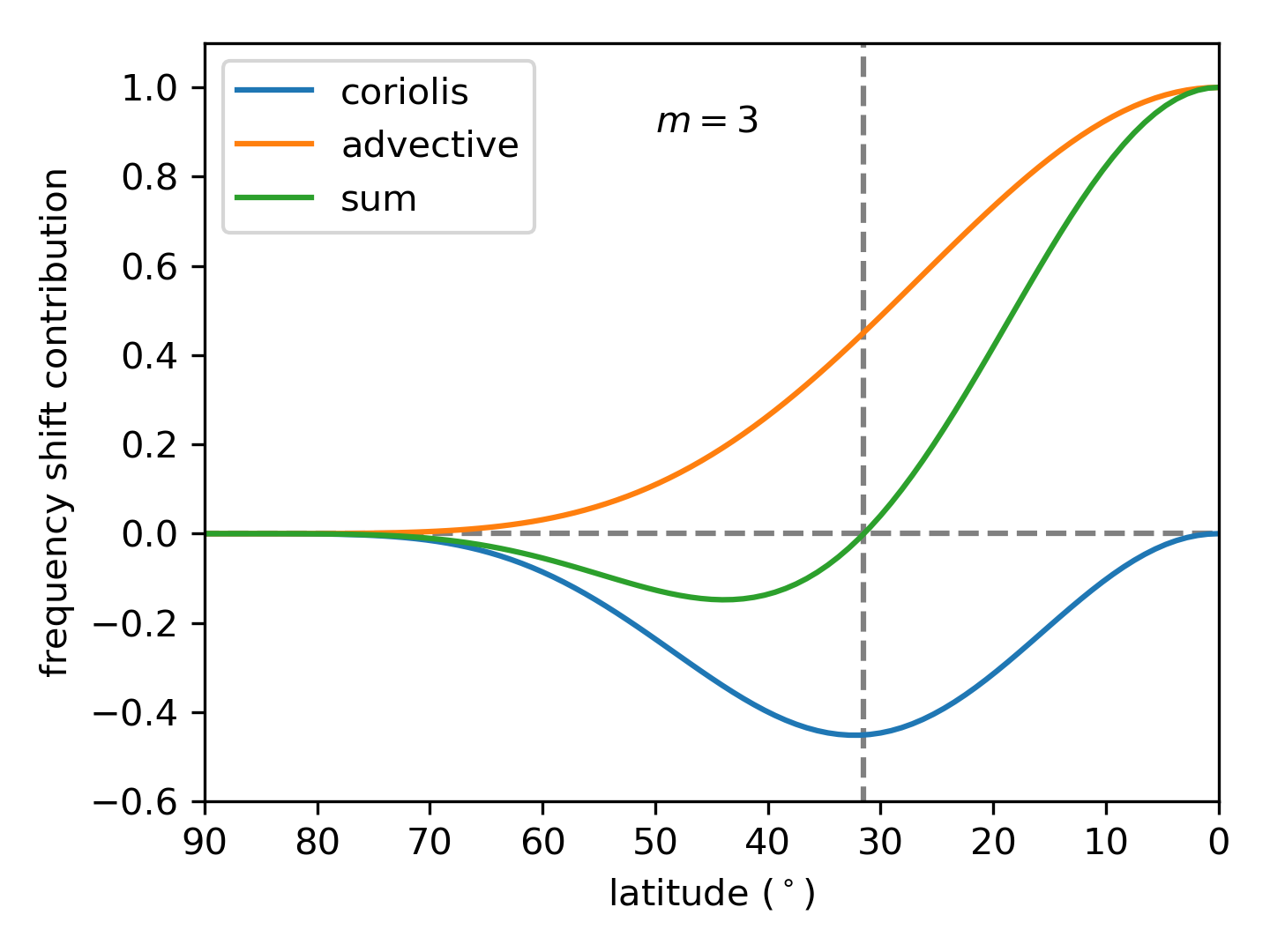}
\caption{Latitudinal dependences of the coriolis and advective terms that contribute to the kernel, and their sum, for $m$ = 3.}
 \label{lat_dep}
\end{figure}

\begin{figure}
\centering
\includegraphics[width=9cm]{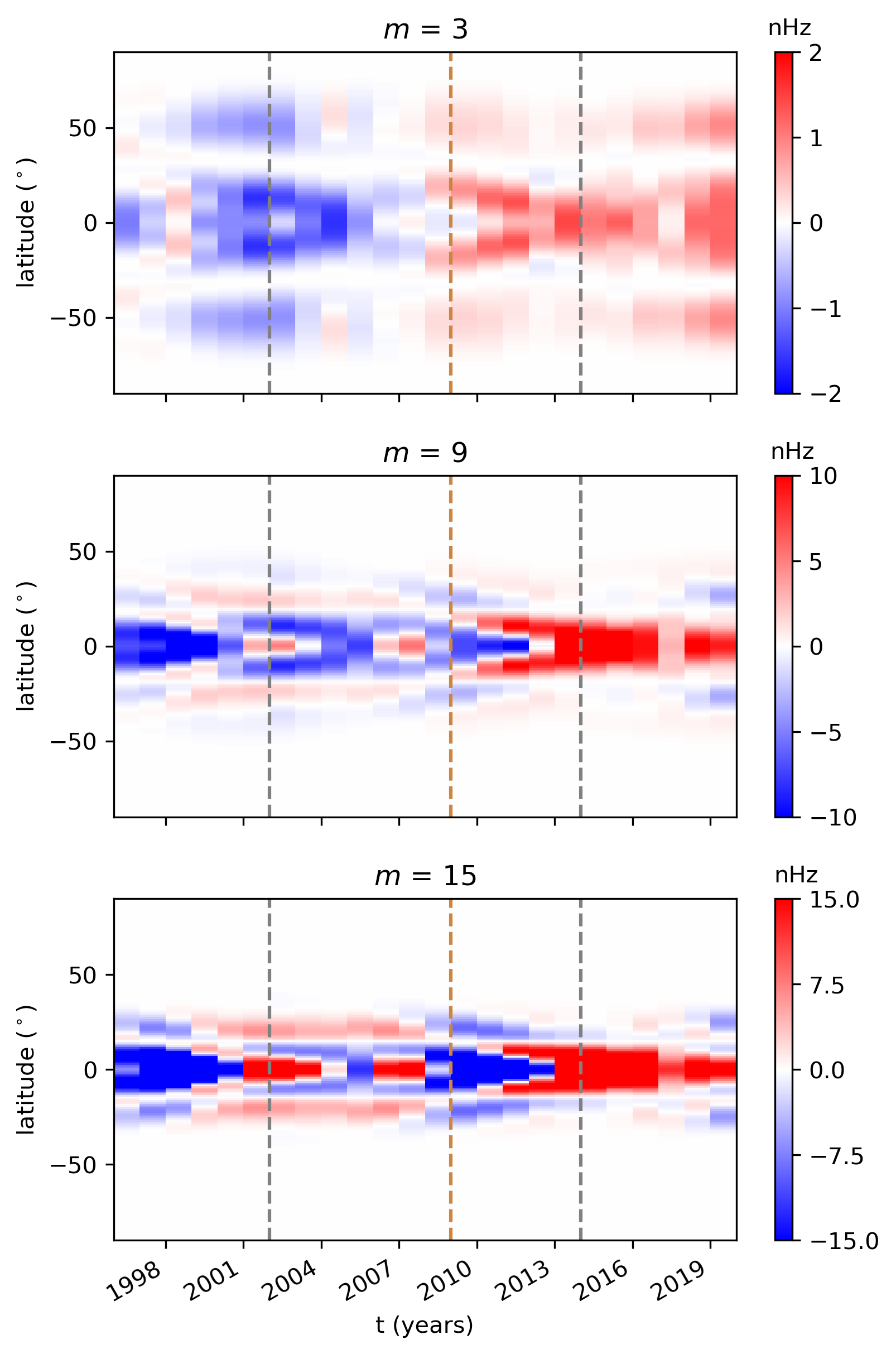}
\includegraphics[width=8cm]{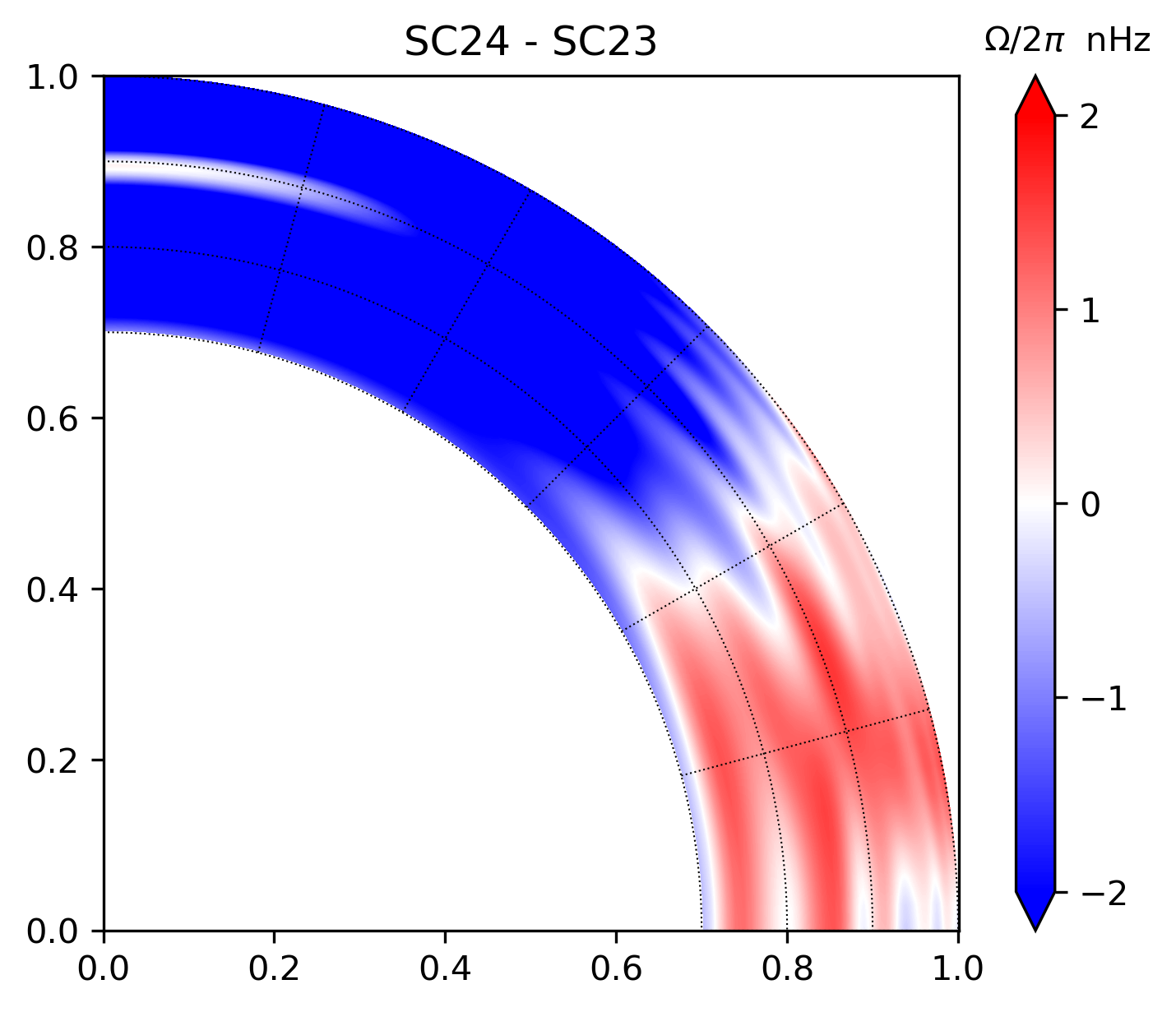}
\caption{\textit{Top three panels:} Rotation rate residual multiplied by the sensitivity kernel and integrated over the radial coordinate.  
\textit{Bottom panel:} Difference in the solar rotation rate in the convection zone between the maxima of cycles 24 and 23.}
 \label{fig:rot_Ker_int}
\end{figure}

The figures in Section~\ref{ker} show that the sign of the kernel switches at a given latitude, which decreases as the azimuthal order increases. This behaviour is due to the terms from the advection and coriolis force in the final kernel having opposite signs, and different latitudinal dependences.  The remaining advective term in the kernel is proportional to ${m(\sin\theta)^{2m-1}}$ and the remaining coriolis term is proportional to ${-(m^2+2) (\cos\theta)^2 (\sin\theta)^{2m-1}}$; therefore, they are equal and opposite at
\begin{equation}
\theta_s = \acos  \sqrt{\frac{m}{m^2 + 2}} .   
\label{eq:theta_s}
\end{equation}
Figure~\ref{lat_dep} shows the latitudinal dependence of the coriolis and advective terms of the kernel, and their sum, for $m$ = 3. The vertical dashed line shows the position as which the two terms are equal and opposite, summing to zero ($\theta_s$). As $m$ increases this position shifts towards the equator and the relative contribution of the coriolis term decreases.

\section{Explaining the  behaviour of  $\delta\omega$ for high and low $m$ values}
\label{app:ker_int}
The top three panels of Fig.~\ref{fig:rot_Ker_int} show the result of multiplication of the rotation rate residual and the kernel, $\delta\diffrot\ K (r,\theta)$, with integration over the radial coordinate. These plots aid in understanding the connection between the rotation data and the calculated frequency shifts for the Rossby waves. We note that faster rotation than the mean profile will result in a positive frequency shift (except where the sensitivity kernel is negative).

For $m$ = 3, we see that both high and low latitudes have a negative contribution to the frequency shift for SC23 and positive contribution for SC 24. This is due to SC23 having lower rotation rates at low latitudes and higher rotation rates at high latitudes compared to SC24, in combination with the kernel being negative for higher latitudes (see Fig.~\ref{ker_plots}). This is shown in the lower panel of Fig.~\ref{fig:rot_Ker_int} where the difference between the rotation residual at the maxima of SC23 and SC24 is plotted. For $m$ = 9 we see a combination of this, and a strong contribution at low latitudes from the zonal flows, which becomes even stronger for $m$ = 15 as the kernel becomes more constrained to low latitudes near the surface (see Fig.~\ref{ker_plots}). In all three plots the switch of the kernel from positive to negative is visible as the pair of horizontal lines of zero frequency at latitudes $\pi/2 - \theta_s$.

\section{Comparison with calculations in the $\beta$-plane }
\label{app:beta}
Here we perform a comparison with recent direct numerical calculations of frequencies of viscous Rossby modes in the equatorial $\beta$-plane 
\citep{2020gizon}
to validate our first-order frequency shifts.
These calculations cannot inform us about the effects of the radial variations in rotation, but they can be performed for any latitudinal variations in the zonal flows.
In addition, the $\beta$-plane approximation is only valid for $m \gg  1$.

To make a direct comparison, we define the zonal velocity in the $\beta$-plane as a function of the coordinate $y=R_\odot  \cos{\theta}$ as
\begin{equation}
    U(y, t) = \frac{\int_0^{R_\odot} r \sin\theta\ [\Omega(r, \theta, t)-\Omega_{\textrm{eq}}(r)]\ r^{2(m+1)} \rho\id r}{\int_0^{R_\odot} r^{2(m+1)} \rho\id r} ,
\end{equation}
where $\Omega_{\textrm{eq}}(r)$ is the equatorial rotation rate as a function of radius. This flow is used as an input to the computations described in \citet{2020gizon}. 
In addition, we specify the viscosity through the Reynolds number  $Re = 300$.

The $\beta$-plane calculation is performed at the full 72-day cadence of the rotation data, and the frequency-shift time series for each $m$ is then mean subtracted to give a result equivalent to our first-order calculations. These are then binned into calendar years. 

\begin{figure}
\centering
\includegraphics[width=9cm]{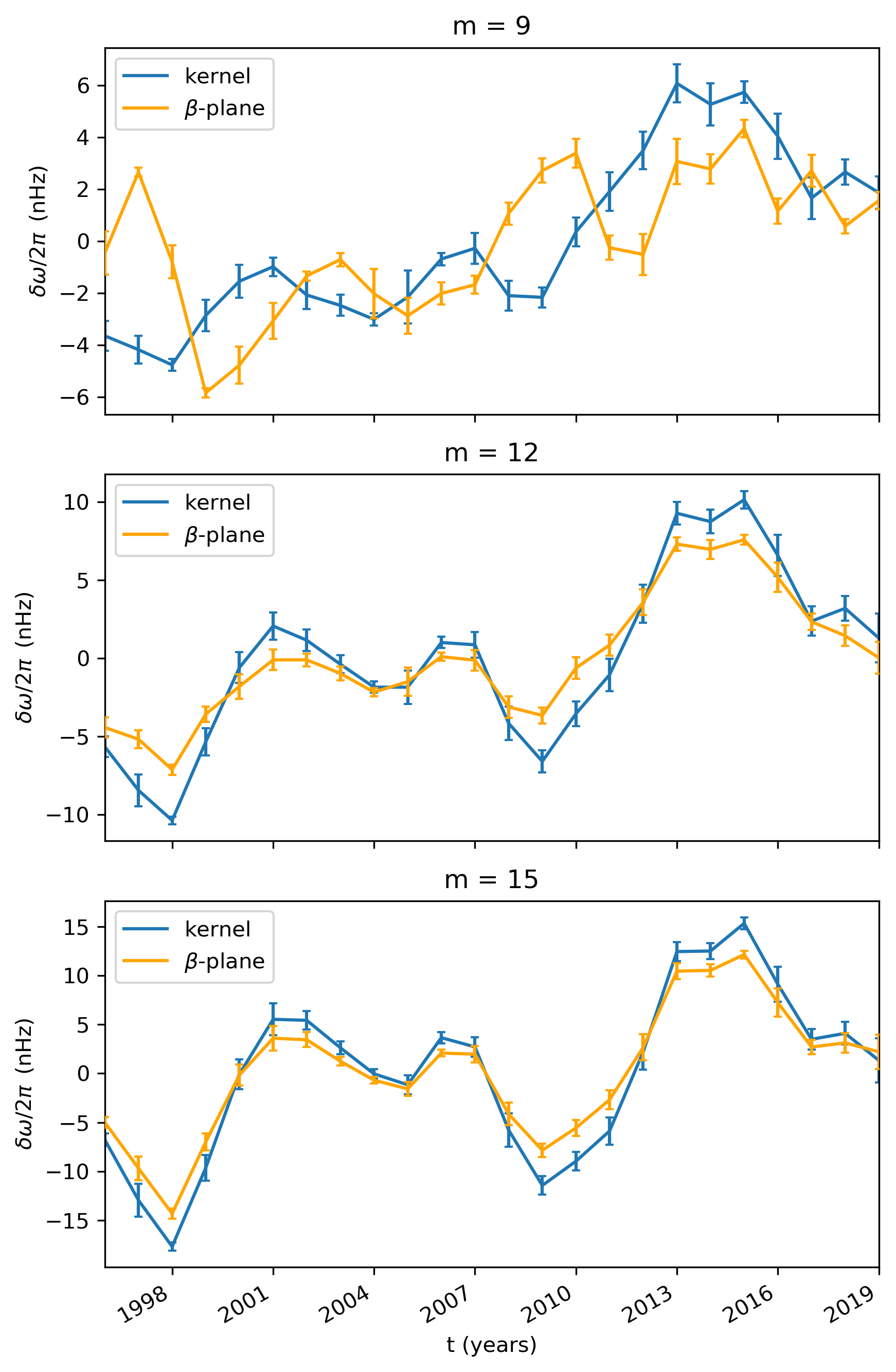}
\caption{Frequency-shift time series from calculations in the $\beta$-plane (orange) plotted along with the results from the kernel (blue) for $m=$9, 12, and 15.}
\label{fig:beta_comp}
\end{figure}

Figure~\ref{fig:beta_comp} shows the resulting frequency-shift time series   along with our first-order results  for $m=$ 9, 12, and 15. There is good agreement for $m=$ 12 and 15. The agreement is still reasonable for $m=9$. For even smaller values of $m$ the agreement becomes worse as the problem is less suited to the $\beta$-plane approximation. 

\end{appendix}

\end{document}